# AI-Driven Cybersecurity Threat Detection: Building Resilient Defense Systems Using Predictive Analytics

**Biswajit Chandra Das [1] *, M. Saif Sartaz [2], Syed Ali Reza [3], Arat Hossain [4], MD Nasiruddin [5], Kanchon Kumar Bishnu [6], Kazi Sharmin Sultana [7], Sadia Sharmeen Shatyi [8], MD Azam Khan [9], Joynal Abed [10]**

[1] *Associate degree in Computer Science, Los Angeles City College*
[2] *Electrical Engineering and Computer Science, Florida Atlantic University*
[3] *Department of Data Analytics, University of the Potomac (UOTP), Washington, USA*
[4] *Information Technology Management, St Francis College*
[5] *Department of Management Science and Quantitative Methods, Gannon University, Erie, PA, USA*
[6] *MS in Computer Science, California State University, Los Angeles*
[7] *MBA in Business Analytics, Gannon University, Erie, PA*
[8] *Master of Architecture, Louisiana State University*
[9] *School of Business, International American University, Los Angeles, California, USA*
[10] *Master of Architecture, Miami University, Oxford, Ohio*
*Corresponding author E-mail: dasbc9672@student.laccd.edu*

## Abstract

This study examines how Artificial Intelligence can aid in identifying and mitigating cyber threats in the U.S. across four key areas: intrusion detection, malware classification, phishing detection, and insider threat analysis. Each of these problems has its quirks, meaning there needs to be different approaches to each, so we matched the models to the shape of the problem. For intrusion detection, catching things like unauthorized access, we tested unsupervised anomaly detection methods. Isolation forests and deep autoencoders both gave us useful sig-nals by picking up odd patterns in network traffic. When it came to malware detection, we leaned on ensemble models like Random Forest and XGBoost, trained on features pulled from files and traffic logs. Phishing was more straightforward. We fed standard classifiers (logistic regression, Random Forest, XGBoost) a mix of email and web-based features. These models handled the task surprisingly well; phishing turned out to be the easiest problem to crack, at least with the data we had. There was a different story. We utilized an LSTM autoencoder to identify behavioral anomalies in user activity logs. It caught every suspicious behavior but flagged a lot of harmless ones too. That kind of model makes sense when the cost of missing a threat is high and you're willing to sift through some noise. What we saw across the board is that performance wasn't about stacking the most complex model. What mattered was how well the model's structure matched the way the data behaved. When signals were strong and obvious, simple models worked fine. But for messier, more subtle threats, we needed some-thing more adaptive, sequence models and anomaly detectors, though they brought their trade-offs. The takeaway here is clear: in cybersecu-rity, context drives the solution. There's no universal model that works for everything. The smart move is to build systems that fit the prob-lem, and more importantly, evolve with it. Threats don't sit still, and neither should our defenses.

## 1. Introduction

### 1.1. Background and motivation

Cybersecurity isn't what it used to be, as the tools that worked a few years ago are struggling to keep up with the pace and complexity of today's industry threats. Malware can now change its form, zero-day exploits are popping up more often, and sometimes the danger comes from inside the network itself. Fixed-rule systems and signature-based detection made sense when most attacks followed a predictable script. That's not the case anymore. These traditional setups now have an unpredictable tendency to unravel when they come upon some-thing unfamiliar. They're good at catching what they already know, but not so great when the threat looks new or doesn't follow the usual patterns. With more companies shifting to the cloud, expanding remote access, and leaning on sprawling supply chains, there are simply more ingress points for attackers to tap into. Almomani et al. (2023) highlight this problem clearly, drawing attention to the fact that signature-based systems frequently fall short when it comes to threats that are disguised or entirely novel, mainly because they were not built to adjust on the fly [2]. This is where AI and machine learning come in, not as some futuristic idea, but as something that's already proving necessary.

Models trained to detect behavior patterns and anomalies can spot threats even if they've never seen them before. Berman et al. (2019) showed that these approaches work even without labeled attack data, something traditional systems can't manage [3]. So, instead of reacting after the fact, security teams can start to anticipate and prevent. There's also a lot happening in more specialized areas. Deep learning models like LSTMs and autoencoders are being used to pick up on small shifts in system behavior that may point to an upcoming attack.. For structured data, methods like XGBoost are especially effective and fast, which is ideal for malware detection, where every second counts. Even in phishing detection, an area long dominated by basic blacklists, we're seeing NLP tools like BERT outperform older techniques by a wide margin. The point is, the nature of cyberattacks has changed. They're not only faster and more complex, they're also showing up in more places: your cloud apps, your endpoints, your internal traffic. It's no longer enough to have detection tools that wait around for something obvious to trigger an alert. Ahmed et al. (2021) point out something straightforward: when teams use predictive analytics, they respond quicker and get breached less often [1]. That's not because they have flashier tools; it's because they're thinking ahead in a landscape that changes by the hour. Security isn't something you set up once and walk away from anymore. It's a moving target. And AI-driven systems are starting to take center stage, not because they're new or exciting, but because they help you keep up. Without that kind of support, you're constantly reacting instead of staying in front of the problem.

### 1.2. Importance of this research

As more organizations shift to digital-first models, their survival increasingly hinges on keeping their systems secure and always available. Whether it's a bank fighting off credential stuffing attacks or a hospital working to protect patient data, the scope of what counts as cybersecurity has expanded. And while budgets for security have grown, defending a sprawl of devices, remote servers, and cloud platforms remains a serious challenge. Most of the old security tools weren't built for this kind of complexity, as they tend to focus on reacting to incidents after the fact, rather than spotting subtle early signs before things go wrong. Worse, they often struggle to scale. The asymmetry doesn't help either; an attacker only has to succeed once, while defenders have to be on point every time. One of the biggest blind spots in traditional systems is their inability to catch early warning signs that aren't obvious. Malware can easily sneak through by concealing its true form. Phishing emails don't need to look overtly malicious; they can also exploit the kind of language quirks that spam filters might miss. And insider threats are especially tricky, since they often operate through legitimate channels. As Stiawan et al. (2023) point out, over 70% of breaches today involve things like stolen credentials or social engineering, and these aren't problems you can solve with signature matching alone [16]. When you dig into it, machine learning really fits this problem's needs. Once you've trained a model well, it learns what normal behavior looks like across users, systems, and networks and notices those subtle deviations that can signal trouble. It catches patterns that rule-based tools often miss, especially when the data is messy and high-dimensional. As more U.S. businesses and public institutions move their workflows into digital spaces, especially in finance and city services, the need for security that's built in from the start becomes harder to ignore. Ray et al. (2025) pointed out that machine learning and AI aren't just useful tools here; they can help stabilize digital finance systems in urban environments. But they also made it clear that for any of that to work, the underlying data pipelines need strong threat detection baked in. Without that, the whole system's at risk [15].

Chio and Freeman (2018) highlight how unsupervised models like autoencoders or isolation forests can detect insider threats simply by noticing changes in behavior, even if you've never seen that exact attack before [6]. And with adversaries constantly evolving their tactics, it helps to have models that can generalize beyond what they've been explicitly trained on. In phishing detection, for instance, word vector models paired with decision trees or logistic regression have outperformed traditional filters by picking up on deeper semantic patterns that rules tend to miss. Then there's the added complexity from cloud-native systems, CI/CD workflows, and edge devices. Security isn't its separate world anymore; it overlaps with behavioral science, systems engineering, and software development. This research leans into that complexity. Instead of treating cybersecurity as one massive problem, it breaks it down into focused areas, intrusion, malware, phishing, and insider abuse, and looks at how different machine learning models fit each one. That kind of modular approach mirrors how security teams work and makes the findings more applicable.

### 1.3. Research objectives and contributions

This research explores how various machine learning and deep learning models handle cyber threats across four distinct areas: spotting intrusions, classifying malware, detecting phishing attempts, and identifying insider threats. We aimed to really understand each model's strengths and weaknesses, and how they behave in situations that mirror real-world challenges. We approached each type of threat individually, always keeping in mind its specific data characteristics and operational needs. For intrusion detection, we focused on unsupervised models, like isolation forests and autoencoders. The idea there was to see if they could flag suspicious activity even without pre-labeled threat examples. When it came to malware, ensemble models such as XGBoost and Random Forest felt like the right fit. They're quite good with the kind of structured, high-dimensional data you get from network traffic or file details. Phishing detection required a slightly different approach. Here, we were dealing with email and website content, so we trained models like logistic regression and XGBoost on feature-rich datasets built to reflect the real-world variations you'd encounter. And for insider threats, which are all about spotting evolving patterns, we chose sequence models, LSTMs specifically. We fed them synthetic user activity data – think logins, access logs, session durations – the kind of information that often signals something amiss internally.

A big part of this whole effort involved building datasets that felt like real-world scenarios, even if they were synthetic. We made sure these datasets truly reflected cybersecurity's messy side: imbalanced data, noisy signals, and behaviors that aren't always neatly 'good' or 'bad'. We also used SHAP. That helped us peek inside the models to grasp which features were pushing their decisions. That kind of insight is vital if you want to trust these tools in a real Security Operations Center. By carefully matching the right models to each threat type and meticulously weighing things like precision, recall, and false positives, our study offers a flexible framework. It's designed to adapt to various security needs. We think this work can be a solid stepping stone for future research and a very practical starting point for any team looking to bring AI into their security operations.

## 2. Literature review

### 2.1. Traditional threat detection systems

For a long time, cybersecurity has leaned heavily on signature-based and heuristic systems. These tools have done a lot of the heavy lifting in defending networks and devices. Signature-based systems, for example, work by spotting known patterns, whether in malware code, suspicious network traffic, or log activity. Antivirus software is a classic example: it checks what it sees against a long list of known threats (Ahmed et al., 2021) [1]. That works well if what you're facing is something that's already been identified. But if the threat is new, or if the code has been cleverly disguised or altered, these systems often miss it (Almomani et al., 2023) [2]. That's because signatures are static. They're built to catch exact matches, not to recognize clever variations or entirely new methods. The same kind of weakness shows up in spam filters or email security rules. Those rely on simple logic and known patterns, so when attackers change the language or formatting, the system often fails to catch the threat.

Heuristic systems try to go a bit further by setting rules or thresholds for what counts as suspicious. For instance, if there are too many failed login attempts, or if a device starts scanning a bunch of ports it normally wouldn't, that might raise a flag. But even here, the system is only as good as the rules it's been given. These thresholds are tricky to set and need constant adjustment. Get them wrong, and you're either buried in false alarms or blind to real threats (Berman et al., 2019) [3]. And let's be honest, security teams are often overwhelmed. Too many false positives, and people stop paying attention. Too many false negatives, and the damage is done before anyone realizes what happened. These systems wait for something to go wrong, then react. They don't anticipate, and they don't adapt well. SIEM platforms and similar tools try to piece together signals from different logs and events. They look for patterns that might mean something's off. But unless someone's already defined the exact pattern to look for, these systems usually don't catch it. They often work within narrow data silos and depend on human-written rules that take time and expertise to get right. And when those rules don't account for how attackers move laterally through a system or blend across domains, critical clues fall through the cracks.

### 2.2. Machine learning in cybersecurity

Machine learning has stepped in to fill some of the big gaps that traditional cybersecurity systems struggle with. It brings flexibility—models can adapt, spot patterns they haven't seen before, and pull useful signals from messy data without having everything spelled out for them. Take malware detection. Decision trees and random forests have been trained on features like opcode frequency, metadata, and how files are structured. These models have had some solid wins, especially when it comes to catching new malware strains that don't match any known signature. Support Vector Machines have been put to work in spam and phishing detection. In those cases, people have hand-crafted features that capture weird word choices, shady domain names, and email timing patterns that don't quite look right. When the data is tabular and the feature space gets large, models like XGBoost have shown they can handle the complexity without choking.

A lot of this progress rides on good datasets. EMBER, for instance, gives you PE-file features to train models that don't rely on simple hash matches. PhishTank and the Nazario dataset pull in real phishing URLs so text-based models can learn to flag suspicious links by looking at token patterns and odd domain combinations. The CERT dataset logs simulated insider threat scenarios, giving researchers labeled user behavior to work with, something you rarely find in the wild. These models have outperformed traditional systems in a number of ways. Some can flag stolen credentials or detect sketchy file access without needing an obvious technical anomaly. But they aren't perfect. They still lean heavily on labeled data, which is hard to come by, especially for insider threats, where real attack examples are rare. That kind of imbalance makes models fragile (Ahmed et al., 2021) [1]. And despite the push toward automation, most systems still need human input to decide which features matter or how to shape them. So far, machine learning has worked well in structured setups and has made decent progress in phishing detection. But it struggles when things get messier, like dealing with logs from identity providers, cloud platforms, or third-party services. Those environments mix formats, data types, and signal quality.

Lately, machine learning has also been showing real promise in transaction monitoring, especially when it comes to spotting subtle forms of fraud, things like duplicate transfers or odd patterns buried in transactions that otherwise look normal. Fariha et al. (2025) showed how combining supervised learning with richer feature sets can noticeably tighten the net around financial fraud. What's interesting is that their approach isn't far off from what we see in malware and phishing detection. The same core ideas, thoughtful feature engineering, and well-tuned models carry over, which makes sense given how these problems often boil down to recognizing patterns that don't quite fit, even when they seem legitimate at first glance [7].

### 2.3. Deep learning and sequential modeling

Deep learning has become more common in cybersecurity because it handles the kinds of patterns that traditional machine learning and rule-based systems tend to miss. Matters involving non-linear relationships and time-based behaviors, patterns that don't fit neatly into simple logic, are where these models start to display utmost competence. Recurrent neural networks, especially LSTMs and GRUs, are good at dealing with sequences; they are able to pick up on the rhythm and structure of events over time, like login patterns, file access sequences, or the order of system commands. Take user behavior analytics as an example. Tuor et al. (2017) showed that LSTM models could learn what's "normal" for each user when it comes to logins or accessing files. Once the model understands that baseline, it can catch small deviations that might hint at insider threats, even when there aren't many labeled examples to train on [18]. Similarly, deep autoencoders have been used to flag unusual patterns in network traffic, like DDoS attacks or lateral movement. The idea is simple: if something's very different from what the model has seen before, it'll struggle to reconstruct it, and that error becomes the signal.

There's also been a move toward hybrid setups that bring together the strengths of different architectures. Hossain et al. (2025), for example, tested out LSTM, Bi-LSTM, and attention-based RNNs on energy usage data and found that these models were great at spotting long-term dependencies and predicting rare events [8]. Even though their work focused on utilities, the structure holds up when applied to cybersecurity, especially for tracking sequences of actions and identifying when things go off-script. In a related space, attention layers paired with CNNs have been used to flag malicious file behavior, first extracting small local patterns and then looking at how they unfold over time. When it comes to detecting malware or phishing, deep learning has pushed things forward by moving away from hand-crafted features. Some researchers have trained convolutional networks on visual representations of executable files, turning binary code into images, and found that these models could classify threats with impressive accuracy.

Others have used language models like BERT to analyze the actual text of emails and URLs, spotting subtle linguistic tricks that phishing attempts often rely on. That shift, from relying on structural cues to focusing on semantic content, has opened new ways to catch attacks that play on language and meaning. That said, the picture isn't all smooth. These models need a lot of diverse, labeled data to train properly, and insider threat datasets are notoriously thin. Even when the models do learn something useful, they can be noisy, and false positives are still a real issue, especially for autoencoders and sequential models. There's also the explainability problem. Security teams don't want to act on a model's output unless they understand why something got flagged. Tools like SHAP help, but there's still a long way to go in making these systems more transparent to the people who rely on them.

### 2.4. Gaps and challenges

Even with all the progress in AI-powered cybersecurity, there are still several challenges that make it hard to neither fully rely on these systems nor roll them out at scale. One of the trickiest problems is class imbalance. The events we care most about, like insider threats or zero-day malware, make up such a tiny slice of all activity that even a low false-positive rate ends up flooding analysts with alerts. Researchers have tried to handle this with resampling techniques like SMOTE, but those come with their tradeoffs. They need careful tuning, and they still fall short when it comes to replicating the kinds of behavior real attackers use. Then there's the issue of labeled data, or more accurately, the lack of it. For example, insider threat detection often relies on understanding what's normal for each user over time. That means you need months of baseline data per person, and even then, the actual malicious events are so rare that supervised models don't have much to learn from. Unsupervised and semi-supervised models help to some degree, but they often miss the context you'd get from richer logs (Ahmed et al., 2021) [1]. In phishing detection, models that rely on tokenizing email text still struggle with subtle tricks attackers use, like context manipulation or semantic wordplay.

Adversarial attacks are another big hurdle. A lot of the more complex models, especially deep learning or ensembles, are surprisingly easy to fool. Malware authors deliberately tweak code to throw off classifiers. Phishing campaigns shift domains and wording constantly to sidestep filters. And even when the models work, it's often hard to explain why. Tools like SHAP or LIME offer some insight after the fact, but building explainability directly into fast, real-time systems is still an ongoing challenge. Models like deep autoencoders and gradient boosters can be evaded through subtle input manipulation. For instance, in malware detection, attackers might inject harmless code blocks to change the ratios of opcodes without affecting the overall functionality. In phishing attacks, replacing legitimate URLs with homographs (such as "google.com") can often bypass standard detection methods. These attack vectors undermine the reliability of static classifiers unless adversarial training or robust feature validation is applied. Also, cybersecurity isn't one big unified problem; it's a bunch of smaller, very different ones. A model trained to catch malware probably won't help with user behavior analytics, and vice versa. There's no universal solution. What we need are flexible, modular systems that can adapt to different data types and threat categories. They should be built with care around issues like imbalance, adversarial noise, and shifting attacker behavior. Our work tries to move in that direction. We build on what's already been done, but add more realistic synthetic data, tailor models to the task at hand, and bring explainability into the full detection pipeline, not as an afterthought, but as part of the design.

## 3. Methodology

### 3.1. Threat domains and data collection

We focused on four key areas in cybersecurity, each tied to a different kind of threat and requiring its own approach to data generation. For intrusion detection, we built a synthetic dataset that mimics a typical enterprise network. It included internal IPs, web servers, and everyday user-to-server interactions. The clean traffic was based on real Firewall and NetFlow stats from open research sources. To simulate attacks, we introduced things like port scans, half-open TCP connections, and traffic bursts, tweaking known attack patterns to make them feel more like stealthy reconnaissance attempts. For malware classification, we created a dataset using file-level metadata. We pulled things like opcode frequency, section entropy, and signature flags from public sources like VirusShare and EMBER. Clean files came from vetted software lists, while malware samples were modified with simulated packing and obfuscation to keep things varied. We also randomized features like file attachment types, import counts, and packer ratios to add realistic noise.

Phishing detection was a bit different. We stitched together a dataset using sanitized Enron emails, PhishTank URLs, and SpamAssassin's public corpus. Each email record included structure-based features (like how many links it had or whether SPF checks passed), keyword signals, and rough estimates of sender reputation from public abuse-report APIs. To make things messy in a realistic way, some legit emails were made to look suspicious (say, weird send times or SPF fails), and some phishing samples looked clean on the surface. Lastly, we tackled user and entity behavior analytics (UEBA). We leaned on DARPA and CERT insider-threat logs to build a base of user session templates, logins, file access, commands, and privilege use. Then we added layers of synthetic behavior across 100 user profiles over 30 days. This included both typical activity and oddities like late-night database queries, multiple failed logins, or risky file transfers. None of these datasets were published as-is. Instead, we used public structures and academic benchmarks as a starting point, then reshaped and corrupted the data in controlled ways to better reflect the noise and unpredictability you'd see in real enterprise settings. Although these datasets provide a structured foundation for evaluating models, they often lack the complexity found in real enterprise environments. Real systems exhibit greater variability in noise levels, asynchronous event timing, and multi-stage attacks, which are not always captured by benchmarks such as UNSW-NB15 or CICIDS2017. For instance, Moustafa and Slay (2015) pointed out that while UNSW-NB15 offers a good balance of attack types, it fails to encompass the full range of Advanced Persistent Threats (APTs) or time-based privilege escalations [12]. Also, while datasets like UNSW-NB15 and CICIDS2017 are foundational, newer benchmarks such as TON_IoT, BoT-IoT, and ADFA-LD offer more comprehensive coverage of edge-device traffic and modern attack vectors. Incorporating these into future iterations of our system would offer greater realism and challenge our models under conditions closer to production-grade traffic.

### 3.2. Exploratory analysis

In the first analysis of the protocol distribution in the intrusion detection dataset (Fig. 1), it's no surprise that TCP takes up most of the traffic. That's expected since most internet activity leans on it, whether it's browsing, sending emails, or moving files around. UDP comes next, which makes sense given its role in things that need low latency, like video streaming or DNS lookups. ICMP shows up in much smaller amounts, mostly for things like network diagnostics and ping requests. What's more interesting, though, is how this breakdown helps us think about security. Because TCP is everywhere, it's also a favorite target for attacks like SYN floods. On the other hand, ICMP's low volume might give attackers cover for ping-based exploits that don't immediately raise red flags. These gaps in protocol volume, what's common and what's rare, can be useful for shaping how we design intrusion detection features. Say you suddenly see a spike in ICMP traffic. That's probably worth a closer look. And then there's the long tail of barely-used protocols. They might seem unimportant at first, but attackers love hiding in the quiet corners of the network where nobody's paying attention. Even though they don't show up much, ignoring them completely could be a mistake.

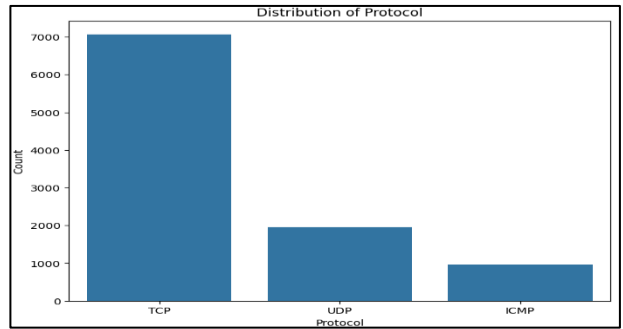

**Fig. 1:** Protocol Distribution in Network Traffic.

Looking at the network traffic (Fig. 2)features gives you a pretty solid sense of how things normally behave and where things start to look suspicious. Source ports tend to be all over the place, which makes sense since they come from a wide range of clients. But destination ports are a different story. They cluster around a few familiar service ports, pointing to standard application use. When traffic starts hitting odd or unexpected destination ports, that's usually worth a closer look. It often signals scanning activity or early stages of an attack. When it comes to how much data is moving and how long connections last, the pattern is lopsided. Most connections are short and light, but there's a long tail where you'll find the occasional big data transfer or lengthy session. Those outliers can mean a few things: maybe someone's pulling data out of the network, launching a denial-of-service attack, or maintaining a hidden channel. Packet count behaves a little differently. It falls into a more typical bell-curve shape, so anything way outside the norm stands out more clearly and could point to abnormal behavior. The is_internal feature tells us that most of the activity stays within the organization's network. When that shifts and more traffic starts heading out, especially to places it doesn't usually go, that can be an early sign something's wrong. Finally, there's the anomaly_label. Like in most intrusion detection datasets, most of the traffic is labeled normal. The rare anomalies are exactly the ones we care about, but their scarcity makes them hard to detect. That imbalance is a real hurdle for machine learning models, and it's something that has to be handled deliberately. Getting familiar with the natural distribution of all these features is the starting point for setting reasonable detection thresholds and catching the things that don't belong.

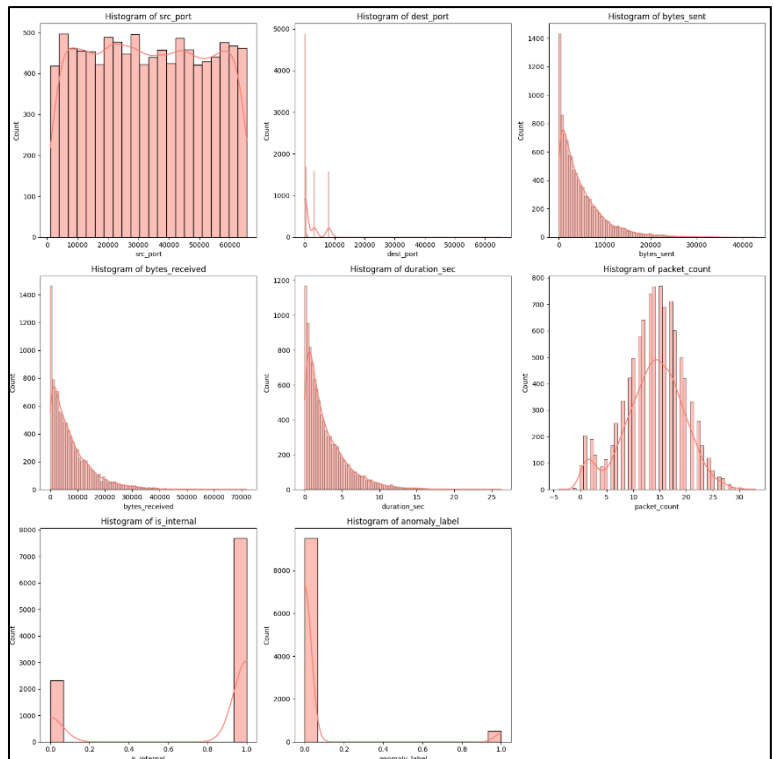

**Fig. 2:** Feature Distribution for Network Intrusion Detection.

When you dig into the file features used for malware detection, some clear patterns start to stand out, patterns that help separate the suspicious from the harmless (Fig. 3). File size and entropy, for example, usually fall into normal distributions. But when a file strays too far from that, it can be a red flag. The same goes for the number of imported libraries and embedded strings. There is a lot of variation, but it's the odd or unexpected imports, things that feel out of place, that tend to hint at something malicious. Opcode ratios like opcode_NOP_ratio and opcode_JMP_ratio generally sit at mid-range values. When they lean too far in one direction, it can signal that the code has been obfuscated or its flow manipulated, both common tricks in malware. One feature that jumps out is has_digital_signature. Most files don't have one, which is already telling, since unsigned files often end up being the shady ones. Then there's section_count. You'll see a few common values pop up across clean files, but when the count veers off the usual path, it might point to techniques like packing or code injection. Speaking of packing, the is_packed feature is heavily tilted toward unpacked files, so when packing does show up, it raises eyebrows. And the packer_entropy_ratio tends to lean to the right, meaning higher values often suggest the file's been purposefully made harder to analyze. Finally, the label feature, which tells you whether a file is benign or malicious, is way off balance. There are far more benign samples than malicious ones, which makes training accurate models tricky. That imbalance can easily skew results unless you take extra steps to handle it.

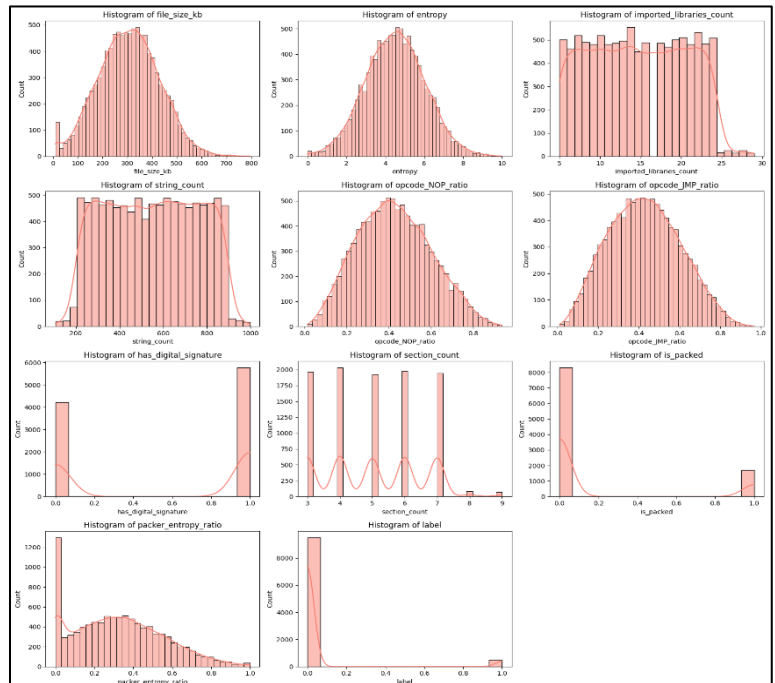

**Fig. 3:** EDA for Malware Data.

When you break down the features behind phishing detection (Fig. 4), a few patterns start to stand out clearly. For example, the has_html feature shows a sharp split. That's not surprising, phishing emails often lean on HTML to mimic real companies or services. Then you've got variables like num_links and num_domains, which are usually low in normal emails. So when you see a message packed with links or pulling from a bunch of different domains, it's worth raising an eyebrow. Some of the strongest signals come from has_spf_fail and is_from_internal. Most legitimate emails pass SPF checks, so when one fails, it's a big red flag; it usually means someone is trying to spoof the sender. And while most real communication comes from within an organization, phishing attempts often come from the outside looking in. The sender_reputation_score also tells a useful story. High scores are common for trustworthy senders, so anything low or dropping fast is suspicious. As for num_suspicious_words, it usually sits at zero. If that number starts climbing, something's likely off. And if an email shows a has_login_form flag, that's a major clue. Benign messages rarely contain login forms, so when they pop up, especially in unexpected places, it's often for the wrong reasons. The hour_sent feature doesn't say much on its own; email traffic is evenly spread out over time. But the big issue is the label feature. There's a serious class imbalance. Legitimate emails massively outnumber phishing ones, which makes training a reliable model harder. Catching the rare malicious ones without overfitting is the real trick.

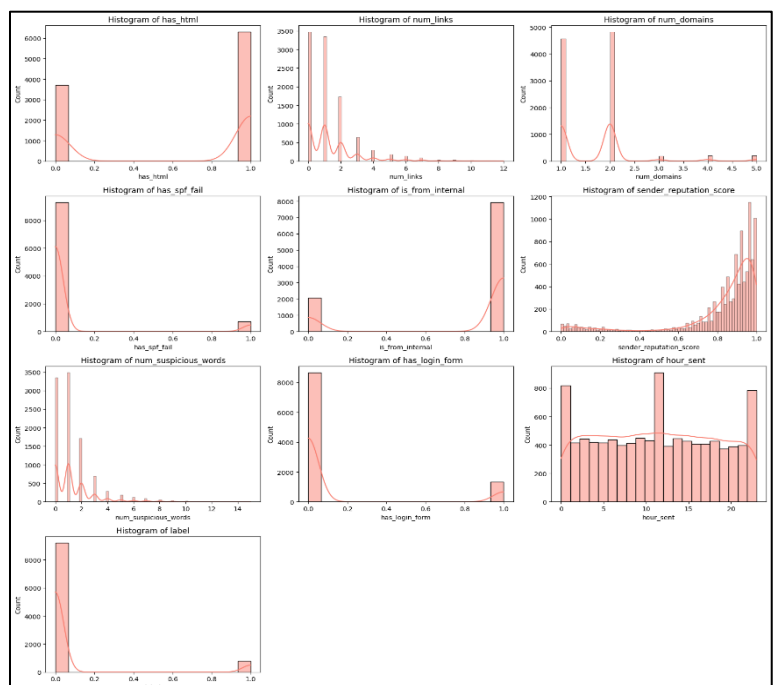

**Fig. 4:** Phishing Email Feature Signals.

Looking at user and entity behavior features gives us a window into how people and systems typically operate (Fig. 5), and more importantly, when something's off. Take the hour feature, for instance. Activity seems to be spread evenly across the full 24-hour range. There are some mild dips and spikes during common work or sleep hours, but nothing too surprising. What matters isn't the overall distribution, though. It's how consistent each user or entity is with their usual behavior. If someone typically logs in mid-morning and suddenly starts accessing systems at 3 a.m., that shift is worth paying attention to. Now, features like accessed_sensitive_file and is_admin_action tell a different story. They're rare, and that's exactly what makes them meaningful. Most users don't access sensitive files or carry out administrative actions as part of their day-to-day work. So when these events do show up, they stand out, and they should. Especially if they don't match the user's usual pattern or occur during odd hours. You see the same kind of imbalance in failed_login_attempts and command_count. Most of the time, users either don't fail logins at all or only fail once or twice. And the number of commands they run tends to be low. But a sudden spike in failed logins? That's a strong indicator that something isn't right, maybe a brute-force attempt or someone poking around where they shouldn't be. A strange flurry of command activity could also suggest someone's using scripts to move around or gather data. Then there's the label, the one that marks whether a behavior is flagged as normal or anomalous. Unsurprisingly, most activity falls into the normal bucket. That heavy imbalance makes it tough to train reliable detection models. They must pick out the rare anomalies buried in a sea of routine behavior without getting tripped up by noise. That's why understanding the baseline, what "normal" really looks like, is so critical. Without that, the subtle shifts that signal real threats are easy to miss.

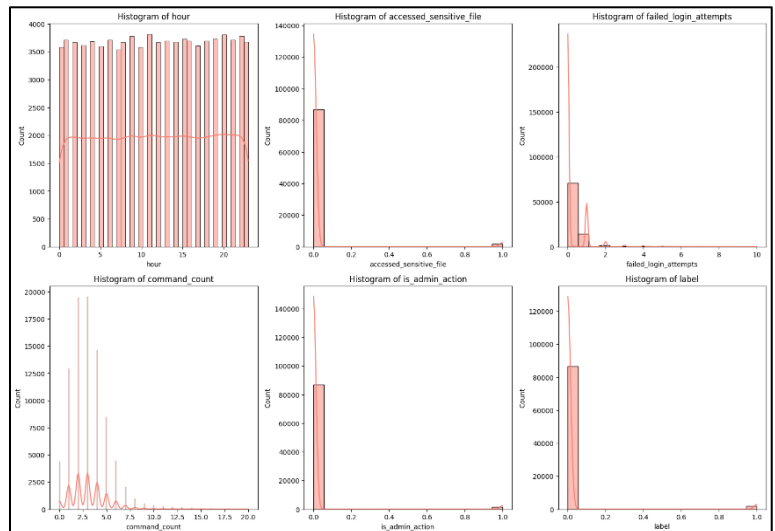

**Fig. 5:** User Session Features in Insider Threat Logs.

### 3.3. Data preprocessing

Across all four domains, we applied a consistent preprocessing framework while incorporating domain-specific steps to prepare data for modeling. First, non-predictive identifiers, timestamps, IP or user IDs, and raw filenames were dropped in every dataset to avoid leakage. For tabular models (intrusion, malware, phishing), categorical features such as protocol types, file extensions, and attachment types were transformed with one-hot encoding (dropping the first level to prevent multicollinearity), while binary indicators (e.g., SPF failure, login form presence) were left as is. Continuous features were scaled with a standard z-score StandardScaler to normalize ranges, except in the malware domain, where tree-based models permitted pass-through of raw counts. Imbalanced classes were addressed differently: intrusion and UEBA used unsupervised anomaly thresholds rather than oversampling; malware employed SMOTE on the training split after encoding; phishing applied random downsampling of the majority (legitimate) class to match the phishing ratio. Splits were stratified 70/30 for all supervised tasks to preserve class proportions.

For UEBA (User Entity and Behavior Analysis) sequence modeling, we engineered additional temporal features: each event log's timestamp was converted to hour and weekday proxies, then grouped into sessions per user per day. After dropping the original timestamp, sessions shorter than a fixed maximum length were zero-padded, and longer sessions were truncated to create uniform 3D tensors of shape (num_sessions, time_steps, num_features). Categorical session attributes, such as activity type, were one-hot encoded with sparse_output=False, while numerical sequence features, such as failed login counts, command rates, and sensitive-file flags, were scaled. The result is an LSTM-compatible array of sequences with corresponding session labels set by the maximum event anomaly flag in each session. By standardizing these preprocessing steps and carefully handling imbalance and sequence structure, we ensured that each model received domain-appropriate, high-quality inputs for training and evaluation.

### 3.4. Model development

For the malware classification task, we leaned on two ensemble models: Random Forest and XGBoost. Both are solid when you're dealing with structured, high-dimensional data, and they give some visibility into which features matter. We one-hot encoded the file-type categories, balanced the dataset using SMOTE (since malicious samples were rare), and then trained a Random Forest with 100 trees. We played around with depth and split criteria to avoid overfitting. At the same time, we trained an XGBoost model and tuned things like learning rate, subsample ratio, and regularization using randomized search. Both models got the same balanced input and were evaluated on separate test sets to keep things honest. Random Forest gave us more stable, low-variance predictions, while XGBoost tried to correct its own mistakes along the way, which gave us a nice contrast in learning behavior.

For phishing detection, we started simple with logistic regression, partly because it's easy to interpret, partly because it sets a good baseline. Then we brought in XGBoost to catch any non-linear patterns that logistic regression might miss. We one-hot encoded attachment types scaled numeric features like link counts and domain age, and downsampled the majority class to get a more balanced view. Logistic regression, with class-weight tweaking, gave us quick probability outputs and clear feature importance; things like SPF failures or fake login forms stood out. XGBoost picked up where it left off, learning interactions on its own. We added early stopping and calibrated probabilities to avoid overfitting and keep outputs meaningful. Both models went through the same pipeline, so we could really see the tradeoffs between transparency and raw predictive performance.

For network anomaly detection, we took an unsupervised route with autoencoders. The idea was to let the model learn what normal traffic looks like, then flag anything that didn't fit. After scaling and one-hot encoding session features, including things like packet rates and byte transfer ratios, we trained the autoencoder only on clean traffic. The architecture was symmetric, with L1 regularization to keep things tight. During inference, we measured how much the model struggled to reconstruct a session. If the reconstruction loss was above a threshold (we used the 95th percentile of the training errors), it was flagged as anomalous. No labeled attacks needed. This setup gives us some flexibility when dealing with new, unknown threats that don't follow familiar patterns.

For insider threat detection (UEBA), we built a sequence-to-sequence LSTM autoencoder. The idea was to model behavior over time, using daily session logs from each user. We encoded each session as a sequence of activity vectors, things like action types, command counts, failed logins, and sensitive file access. The LSTM encoder compressed the sequence into a compact latent vector, and the decoder tried to reconstruct the original pattern. If it couldn't, we flagged it. High reconstruction error meant the session didn't look like what that user usually does. Since we didn't have tons of labeled insider attacks, this approach let us detect shifts in behavior without needing to hand-tag every possible malicious action. Hyperparameter tuning was also applied to improve the model performances for some of the models, and cross-validation was employed to ensure the models do not overfit.

## 4. Evaluation and results

### 4.1. Intrusion detection

We looked at how well two models, Isolation Forest and a deep autoencoder, could detect unauthorized access in a network (Fig. 6). Both hit 93% accuracy overall, which tells us they're decent at telling normal traffic apart from anomalies. But when we dig into the details, especially the rare cases of actual intrusions, the differences start to show. The Isolation Forest had an F1-score of 0.44 for anomalies, with a precision of 0.37 and a recall of 0.55. That means it caught just over half of the real threats, but also raised a lot of false alarms. The autoencoder did a bit better, with an F1-score of 0.48, slightly higher precision at 0.39, and recall at 0.61. Its ROC-AUC was 0.96, which speaks to how well it ranks threats overall. The confusion matrices back this up: the autoencoder caught 304 true positives compared to the Isolation Forest's 224, though it also made more noise with false positives. In short, the autoencoder comes out ahead, but not by a landslide. Both models still have a hard time picking out rare attacks without letting too many false flags through.

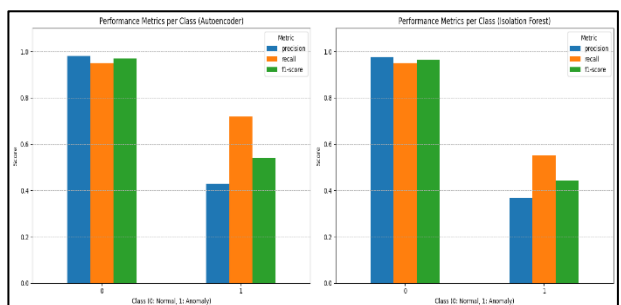

**Fig. 6:** Comparison of Intrusion Detection Model Performance.

### 4.2. Malware detection

For this task, we tested Random Forest and XGBoost on a synthetic file metadata set, trying to sort out benign files from malware (Fig. 7). Accuracy-wise, both hit 97%, which sounds great. But once you look past that top-line number, there's more going on. Random Forest hit a malware precision of 0.80, but recall was only 0.49. So, it correctly flagged most of what it guessed was malware, but missed more than half of the actual malicious files, as the F1-score came out to 0.61. XGBoost did better, with a precision of 0.87, a recall of 0.57, and an F1-score of 0.69, but still fell short of reliably catching everything dangerous. The takeaway here is that both models look solid on paper because benign variables dominate the dataset. But when it trickles down to catching threats, especially the ones that matter, they leave too much on the table. Better recall is going to require either deeper tuning or a smarter ensemble approach.

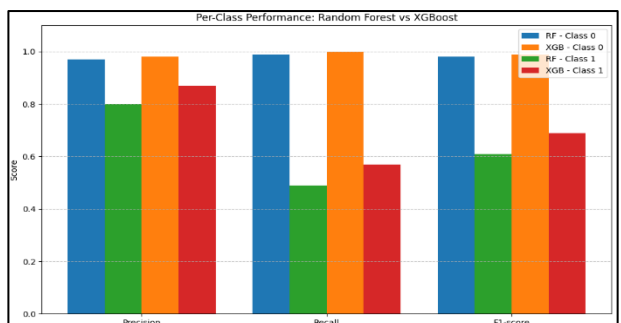

**Fig. 7:** Malware Classification Metrics (Random Forest vs XGBoost).

### 4.3. Phishing detection

This one was almost too easy. Logistic regression, Random Forest, and XGBoost were all trained on structured email and URL features (Fig. 8). Across the board, performance was close to perfect: accuracy between 99.9% and 100%, F1-scores at or near 1.0, and barely any misclassifications out of 3,000 test samples. ROC-AUC values were all right at the ceiling. On the surface, it looks like phishing detection is a solved problem, at least with this dataset. But honestly, results this clean should raise eyebrows. The models may be overfitting or benefiting from artifacts in the data. In a real-world setting, especially with evolving attacks, the story could be very different. While the phishing detection models achieved near-perfect metrics, this likely reflects the cleanliness and separability of the dataset rather than true generalization capability. Real-world phishing attempts are dynamic and often leverage adversarial linguistics, context manipulation, and evolving URL obfuscation, patterns which datasets like Enron or PhishTank may not capture well. As highlighted by Stiawan et al. (2023), modern phishing campaigns increasingly evade static heuristics by mimicking behavioral and semantic traits of legitimate communication [16]. Future evaluations should incorporate more adversarial or zero-day phishing content to assess model robustness better.

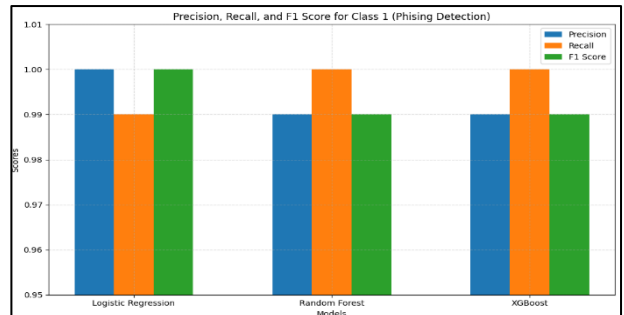

**Fig. 8:** Phishing Detection Model Scores.

### 4.4. User & entity behavior analytics (UEBA)

To detect insider threats, we trained an LSTM autoencoder on sequential user session logs (Fig. 9). It reached 95.7% accuracy and a perfect recall of 1.00 on the threat class, meaning it didn't miss any malicious sessions. That's great on paper. But the precision for threats was only 0.54, leading to a lot of false positives. Its F1-score was 0.70, and the macro-average F1 was 0.84, which reflects balanced treatment of both classes despite the imbalance in the data. If you're working in a setting where missing a threat isn't an option, like a bank or a critical infrastructure system, this kind of model might be worth the noise. But you'll need some post-processing or smarter thresholds to deal with all the alarms it's going to trigger.

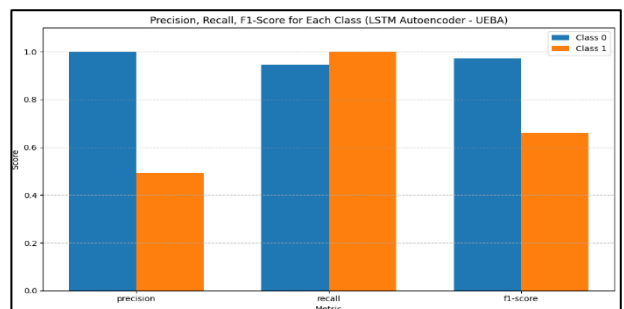

**Fig. 9:** Insider Threat Detection Performance (LSTM Autoencoder).

### 4.5. Reflections across domains

There's no one-size-fits-all model here. Phishing detection looks almost too clean, probably a function of the dataset rather than model brilliance. Malware and intrusion detection are more challenging, especially when it comes to recall on rare events. And the UEBA (User & Entity Behavior Analytics) approach nails sensitivity but struggles with precision, which can become a headache in practice. If there's one consistent thread, it's that accuracy alone doesn't tell the full story. Each domain brings different trade-offs, and finding the right balance between catching threats and minimizing noise is going to depend on what kind of risk you're willing to tolerate.

## 5. Interpretation and operational insights

### 5.1. Interpretation of findings

Looking across four key areas of cybersecurity, the results from evaluating AI-based threat detection models point to a pretty clear takeaway: there's real potential here to build smarter, more adaptive defenses that hold up in the chaos of live operational environments. Start with intrusion detection. The autoencoder pulled slightly ahead of the Isolation Forest, which might seem minor at first, but it matters. Deep learning models like autoencoders are better at picking up on the weird little deviations in network traffic that don't quite trip the usual alarms. That means a security team can catch things like quiet port scans or data exfiltration attempts before they turn into something much worse. Since these models learn what "normal" looks like over time, they don't need constant hand-holding or static rules that go stale. That kind of flexibility makes a real difference in systems that never sit still.

When it comes to malware classification, XGBoost delivered a stronger recall than Random Forest. That tells us it's better at catching sneaky, obfuscated binaries, malware that's trying not to look like malware. In practical terms, endpoint protection tools powered by models like this can stop more malicious files before they get a foothold. Plus, the precision was solid, which cuts down on false alarms that frustrate users and flood IT with unnecessary support tickets. And because these models make it easier to see which features mattered, things like the frequency of jump operations or how packed the binary is, security analysts can respond faster and build smarter rules for next time. Phishing detection was interesting too. Even basic logistic regression models nailed it, thanks to the strong signals baked into email and URL metadata. Features like failed SPF checks or sketchy login forms do a lot of the heavy lifting. The fact that simple models perform this well is a win; it means companies can run them right at the mail server level and stop most phishing attempts cold, without eating up a ton of resources. More complex models like XGBoost help at the margins, so a hybrid setup might make sense: let logistic regression filter the obvious patterns and bring in boosted trees for the trickier edge cases.

Insider threats, though, are a whole different beast. Here, the LSTM autoencoder stood out with perfect recall. It's the kind of model that learns what's normal for each user and flags things that don't fit, like odd late-night database queries or strange command patterns. For organizations, that's a powerful tool, especially since these kinds of threats are subtle and often don't raise any immediate red flags. Yes, the model can be a bit noisy, but ranking alerts by severity or context helps teams focus on what really matters. Stepping back, what ties all this together is the shift from reactive defenses to something more proactive, models that look for behavior, not just known patterns. Jakir et al. (2023) make a solid point: in high-stakes areas like finance or critical infrastructure, it's not enough for a model to be accurate. What matters is whether people can understand how it works and whether it runs fast enough to be useful in real-world conditions. That thinking fits closely with what we saw in this study, especially with the phishing classifiers and UEBA models. They managed to catch threats effectively without slowing things down, which is exactly the kind of balance you need when decisions have to be made in real time [10]. When you combine the strengths of ensemble models, deep learning, and sequence-based approaches, you get tools that help security teams stay ahead of threats, not scramble after them. As shown by Islam et al. (2024), weaving machine learning into SOC workflows isn't about replacing analysts; it's about giving them better eyes and faster reflexes. And in a field where every minute counts, that's no small thing [9]. As cyberattacks evolve, so must model resilience. One forward-looking strategy is adversarial retraining, where simulated evasion samples are used to harden detection models. This allows the system to anticipate and adapt to malicious input perturbations that seek to exploit decision boundaries.

### 5.2. Explainability and trust in AI security systems

If you're building AI models for cybersecurity, explainability isn't optional; it's what makes the system usable. Tools like SHAP and good old feature importance scores go a long way in helping people trust what the model is doing. For ensemble models like Random Forest and XGBoost, feature importance tells you which inputs played the biggest role in a prediction. That might be something like opcode_JMP_ratio for malware detection or num_suspicious_words in a phishing email. Visualizing these helps security analysts quickly check if the model is picking up on real patterns or just latching onto noise. This kind of transparency isn't just for peace of mind; it speeds things up. When an alert comes in, the analyst can immediately see which features tipped the model toward a decision. That means they're not fumbling in the dark. They can jump straight to the email domain or dig into a sketchy file section, depending on what triggered the alert. SHAP adds another layer by explaining individual predictions. Instead of saying "in general, failed logins matter," SHAP shows exactly how much a spike in failed logins pushed a specific prediction toward "malicious." In user behavior analytics, that matters. It means you're not chasing ghosts; you're seeing exactly what drove a session to get flagged. Maybe it was the odd login time or access to a sensitive folder, not some background noise like routine file access. That's gold for security teams trying to focus their time where it counts. It also helps with communication. When your model highlights something intuitive, like an SPF failure or an unusually high packer entropy, it's easier for non-technical folks to follow along. CIOs, compliance officers, risk managers, and people who need to sign off on automated actions are more likely to get on board when the system's decisions make sense. Studies back this up: Buiya et al. (2024) showed that in IoT security, exposing feature-level explanations didn't just improve detection; it helped engineers fix the actual problems faster [5]. Same with fraud detection, Rahman et al. (2024) found that clear explanations helped cut down manual review time and built trust in the alerts [13]. There's also the regulatory side. Standards like NIST SP 800-53 and GDPR are starting to require explainability in automated decision systems. If you're logging SHAP values or feature importances right into your SIEM or UEBA, you're not just helping analysts; you're putting your system in a stronger position for audits and compliance reviews.

In operational environments like Security Operations Centers (SOCs), the need for explainability is crucial for time-sensitive decision-making. SHAP values can be precomputed for common alerts and integrated into dashboards or incident tickets, enabling analysts to quickly identify which features (e.g., SPF failure or opcode_JMP_ratio) triggered an alert. Buiya et al. (2024) highlight that incorporating model insights within Security Information and Event Management (SIEM) tools significantly enhances triage efficiency. This allows response teams to sidestep irrelevant noise and concentrate on high-impact anomalies. Instead of reviewing every prediction, SOCs can rank alerts based on SHAP impact to prioritize their investigation workflows.

### 5.3. Practical implications for U.S. enterprise security

Today's SIEM (Security Information and Event Management) tools, like Splunk, QRadar, or Elastic, are no longer just log aggregators. When you embed models like autoencoders or Isolation Forests into the data pipeline, SIEMs become active participants in threat detection. For example, if an autoencoder spots a weird spike in network traffic that doesn't match the normal pattern, it can flag that in real time. Researchers like Islam et al. (2024) show that plugging these kinds of models into enterprise networks makes a real difference [9]. When systems start looking at user behavior, timing, and context, not just event logs, you get smarter alerts and better visibility into zero-day threats or lateral movements that used to slip by unnoticed. Endpoints are another place where AI can pull its weight. Classic antivirus still leans heavily on signatures, which means it often lags. But with models like XGBoost or Random Forests analyzing file structure and opcode patterns, you can catch new malware variants without needing an update every few hours. You can deploy these models inside lightweight agents, which means they don't need constant cloud calls to make decisions. They can act locally, fast. And in environments built around Zero Trust, where nobody, internal or external, is trusted by default, that kind of responsiveness is essential.

LSTM autoencoders trained on user behavior patterns can monitor sessions as they unfold, flagging actions that don't fit the profile. Berman et al. (2019) highlight how this helps detect insider threats that would otherwise blend in [3]. Rolling out modular AI systems across a distributed security setup, especially in multi-tenant environments, comes with a specific challenge: everything needs to run smoothly across multiple nodes without lag. That's even more important in fast-paced industries like finance or logistics, where even small delays can hurt security performance. Billah et al. (2024) make a solid point here. They show how tuning performance in multi-machine systems can really boost both speed and scalability when handling real-time analytics. That insight carries over directly to large-scale cybersecurity work, where timing and coordination across machines aren't optional; they're essential [4].

Email is still one of the most common ways attackers get through. And while filters have improved, many still rely too much on static heuristics. AI modules, whether it's a simple logistic regression model or something heavier like XGBoost, can catch more by looking at signals like broken SPF records, shady links, odd sender domains, or sketchy wording. The key is keeping false positives low enough that your SOC team doesn't get buried. That's where these models help. Jakir et al. (2023) found that financial institutions using ML-based email filtering saw big drops in business email compromise rates. Attackers don't wait around, and defenses can't either. AI systems that can learn in real time, or adapt based on feedback, give you an edge [10]. With a modular setup, you can swap out or tune models as new

threats emerge, without tearing apart the whole infrastructure. False positives used to be the price of anomaly detection. But that's changing. With ensemble models, probabilistic scoring, and signals from other systems, detection can be smarter and more precise. Buiya et al. (2024) even showed that multi-model systems cut false alarm rates by 40% in some IoT setups [5].

The point of Zero Trust is that you never assume safety; every action must earn it. But pulling that off requires systems that are constantly asking: Does this user behavior make sense right now? Is this device healthy? Does this session feel off in some way? AI models can help make those decisions without grinding productivity to a halt. Deep sequence models, explainable trees, and lightweight classifiers can monitor behavior and score risk as things happen. Rahman et al. (2025) even point to systems that combine blockchain with AI to track behavior securely across distributed environments [14]. As more companies move toward energy-efficient setups for running AI models continuously, especially at the edge, security systems need to keep up. Sultana et al. (2025) suggest using decentralized AI frameworks in edge environments to cut down on energy use without sacrificing performance. Tying this kind of approach into modular threat detection systems could make it easier to deploy security tools across a company's network in a way that's both effective and energy-conscious [17].

### 5.4. Limitations

One of the more obvious sticking points is the kind of data we're using. A good chunk of the datasets in play, like UNSW-NB15 for intrusion detection or the synthetic ones used for phishing and insider threat detection, don't fully capture the messiness of real enterprise systems. They're useful for controlled testing and for keeping class distributions balanced, but they tend to miss the noise, feature drift, and quirks that show up in actual production environments. Moustafa et al. (2015) highlighted that even solid datasets like UNSW-NB15 can end up overweighting some attack types while barely touching on more complex, multi-stage ones [12]. There's also the issue of cleanliness. Pre-labeled datasets are typically scrubbed of ambiguity and borderline cases, which are exactly the types of scenarios that challenge models in the wild. Furthermore, the controlled balance and engineered clarity in public datasets underrepresents the natural entropy and session overlap seen in real enterprise logs, where even benign anomalies may resemble threats, requiring robust contextual understanding. So while models might look great on paper, they often need testing against live traffic, with real-time feedback loops, before they're ready for production. The LSTM autoencoder, for instance, assumes that every session fits neatly into a fixed-length input. That's tidy for training, but doesn't reflect the range of activity we see in practice. Some sessions are over in a flash, others span hours. Padding or chopping them down risks throwing away useful information or watering it down. There are more adaptive approaches out there, Transformers or even dynamic time warping, that might handle this better and are worth exploring. Another challenge is that fixed-length models tend to generalize poorly across different types of users. A senior engineer's daily terminal activity won't look anything like a salesperson's browser patterns. Train one LSTM on the average session, and you risk tuning it to the most common behaviors, ignoring the edge cases that may matter most. Tuor et al. (2017) posit that, for models that take user roles or personalized baselines into account [18].

Then there's the question of thresholds. Autoencoders work by flagging events where the reconstruction error crosses a certain line, but where that line sits isn't obvious. Set it too low and you flood your analysts with false positives. Set it too high and you miss actual threats. Getting it right takes domain knowledge and examples of what "real" anomalies look like, which are often hard to come by. Berman et al. (2019) warned that without regular tuning, even high-performing models can drift and become useless [3]. Changes in software, attacker behavior, or even day-to-day operations can throw them off. One possible way forward is to move toward dynamic thresholds or active learning systems that learn from the alerts they trigger and refine themselves over time. There's also a cost to running these systems at scale. Tools like SHAP or deep sequence models aren't cheap when you're applying them across thousands of devices or emails. The more complex the model, the more infrastructure it demands. That doesn't mean we throw them out; it just means we need to be smart about how we use them. Maybe lightweight models handle the first pass, and more intensive ones get triggered only when something looks suspicious. Finally, controlled lab results are nice, but they don't tell you what happens when these models hit a live system. You don't know how well they'll hold up until they're dealing with real traffic, changing behavior, and attackers trying to work around them. That's where you learn what your false positive rate is, whether your users trust the alerts, and whether the model can keep up with the pace of change.

## 6. Future work

### 6.1. Deployment on real infrastructure in the U.S.

The next big move is to take these models out of the lab and plug them into real enterprise systems. Until now, we've worked with public and simulated datasets, UNSW-NB15, CICIDS2017, and DARPA's Insider Threat data, but those only scratch the surface of the mess and unpredictability you see in a live environment. To see how well our approach holds up, we need to feed it real telemetry: syslogs, firewall alerts, endpoint detection records, and the like. Imagine tapping into tools such as Zeek, Suricata, or Cuckoo Sandbox to gather rich, labeled data from actual network traffic and sandboxed malware runs. You'd get all the extra details, network flows, HTTP requests, DNS lookups, and process activity that help you build time-aware and behavior-focused models. Hooking this into a SIEM gives you streaming alerts and even lets you retrain your models on the fly. Rolling out a small pilot in a U.S. company, say in finance, healthcare, or defense, would show us where these models shine or need tweaking to fit industry-specific threat patterns.

A logical next step is to deploy these AI models in controlled pilots within high-risk sectors. For example, in healthcare, anomaly detection models could monitor EMR access logs to detect unauthorized access to patient records. In finance, phishing detection classifiers could be layered on top of existing fraud monitoring systems for internal email filtering. These pilots would allow us to track false positive rates, response time, and integration cost in domain-specific environments. A logical next step is to implement these AI models in controlled pilot programs within high-risk sectors. For instance, in healthcare, anomaly detection models could be used to monitor electronic medical record (EMR) access logs to identify unauthorized access to patient records. In the finance sector, phishing detection classifiers could be integrated with existing fraud monitoring systems to enhance internal email filtering. These pilot programs would enable us to track false positive rates, response times, and integration costs in specific domain environments.

### 6.2. Graph-based modeling for UEBA and access patterns

Another direction worth exploring is using graph-based models to improve UEBA, User and Entity Behavior Analytics. Most current approaches still treat user activity as flat sequences or rows in a table, which undersells the complexity of how people move through a system. Insider threats don't always show up as obvious spikes or outliers. They often play out as strange paths through a network, accessing systems in an unusual order, escalating privileges in ways that don't fit the usual pattern. A static table won't catch that, but a graph might. By building graphs where users, files, devices, and systems are nodes, and access or interaction is represented by edges, you can start to model not just what happened, but how things are connected. Graph Neural Networks (GNNs) can learn from these structures in ways that traditional models can't. You can also layer in time, using timestamps on edges, so it's not just "who accessed what," but "who accessed what and when."

An easy place to start is by generating graph datasets from existing corpora like the DARPA or CERT datasets. Then the goal would be to eventually transition to logs from live environments, which would give a much richer and more realistic picture. In access-heavy environments like smart grids or enterprise-level energy systems, the connections between users and resources can get tangled. Graph-based models help make sense of that complexity. Instead of treating every user action in isolation, they let us map out the relationships, who's doing what, when, and with which systems. Khan et al. (2025) took that idea further. They combined graph neural networks with fraud detection logic to dig deep into user-resource interactions. Their approach gave them a detailed, layered view of behavior that traditional models often miss. That kind of work sets a strong example for how GNNs could be used to detect insider threats in user and entity behavior analytics (UEBA), where subtle patterns can mean everything [11].

### 6.3. Continual learning and concept drift adaptation

Security setups never stay the same; new weak spots pop up, attackers try out new tricks, and we're always tweaking configurations. Models that stay put end up outdated before long. That means we need a way to keep them learning over time and notice when things shift under our feet. One option is online learning, where the model absorbs fresh labeled data bit by bit instead of waiting for a big overhaul. Approaches like sliding-window updates, sampling techniques that keep a snapshot of past data, or weighting recent examples more heavily help keep the workload manageable. On top of that, drift-detection tools, such as DDM, ADWIN, or Page-Hinkley, raise a flag when assumptions slide out of alignment, and it's time to retrain or fall back to a safe mode. We can even make the system tougher against evasion attempts by feeding misclassified or deliberately tricky samples back into training. Pairing these adaptive methods with explainability tools like SHAP or straightforward feature-importance checks not only makes our defenses stronger but also helps build trust when we're guarding critical systems or handling financial data. Taking these steps moves us closer to a threat-detection setup that scales, stays transparent, and can react on the fly in fast-changing, hostile environments.

## 7. Conclusion

We put together a flexible AI setup that tackles four big areas of cyber defense in the U.S.: spotting network intrusions, classifying malware, flagging phishing attempts, and analyzing user and entity behavior. For each area, we picked models that fit its quirks: Isolation Forests and Autoencoders for network oddities, gradient boosting and logistic regression for phishing and malware, plus LSTM Autoencoders to catch insider shenanigans. Here's what stood out: success came down to matching the model to the problem. No single algorithm ran the table. Phishing detection shone when fed rich email and site features, while our LSTM-based UEBA system nailed every insider-threat scenario we threw at it. In those anomaly-heavy zones, intrusion and malware, deep Autoencoders and XGBoost gave us broader coverage than old-school methods, though we still had to balance recall and false alarms. How we prepped the data mattered a lot. Categorical fields got structured encoding. We used SMOTE or downsampling to even out class skews. We built time-based sequences to help behavior models pick up on patterns. Those steps turned raw logs into the kind of signal our AI could learn from. We worked mostly with synthetic or augmented sets to benchmark our ideas, but we kept things as real as possible. Every piece of this system fits together in modules that can plug into SIEM tools, endpoint guards, or behavior monitors without tearing everything apart. Looking ahead, I'm keen on layering in graph-based reasoning for access links, rolling out online learning so the models evolve with shifting threats, and testing them on live telemetry. As attackers get craftier, our defenses need to match that pace, learning on the fly and explaining their moves to security teams. In the end, cyber defense isn't about chasing a mythical perfect model. It's about building an agile, transparent ecosystem, one where each tool knows its role and shares the load across a threat landscape that never stops changing.

## References

[1] Ahmed, I., Mahmood, A. N., & Hu, J. (2021). A survey of network anomaly detection techniques. *Journal of Network and Computer Applications*, 136, 1–23.
[2] Almomani, A., Al-Ameen, M. N., & Florea, A. M. (2023). A survey of signature-based intrusion detection systems: limitations and future directions. *ACM Computing Surveys*, 56(3), 1–38.
[3] Berman, F., Juels, A., & Westhoff, D. (2019). Insider threat detection in enterprise systems via behavior analytics. *IEEE Security & Privacy*, 17(2), 68–75.
[4] Billah, M., Shatyi, S. S., Sadnan, G. A., Hasnain, K. N., Abed, J., Begum, M., & Sultana, K. S. (2024). Performance Optimization in Multi-Machine Blockchain Systems: A Comprehensive Benchmarking Analysis. *Journal of Business and Management Studies*, 6(6), 357–375. https://doi.org/10.32996/jbms.2024.6.6.18.
[5] Buiya, M. R., Laskar, A. N., Islam, M. R., Sawalmeh, S. K. S., Roy, M. S. R. C., Roy, R. E. R. S., & Sumsuzoha, M. (2024). Detecting IoT Cyberattacks: Advanced Machine Learning Models for Enhanced Security in Network Traffic. *Journal of Computer Science and Technology Studies*, 6(4), 142–152. https://doi.org/10.32996/jcsts.2024.6.4.16.
[6] Chio, C., & Freeman, D. (2018). *Machine Learning and Security: Protecting Systems with Data and Algorithms*. O'Reilly Media.
[7] Fariha, N., Khan, M. N. M., Hossain, M. I., Reza, S. A., Bortty, J. C., Sultana, K. S., ... & Begum, M. (2025). Advanced fraud detection using machine learning models: enhancing financial transaction security. *arXiv preprint arXiv:2506.10842*. https://doi.org/10.14419/c73kcb17.
[8] Hossain, S., Miah, M. N. I., Rana, M. S., Hossain, M. S., Bhowmik, P. K., & Rahman, M. K. (2025). Analyzing trends and determinants of leading causes of death in the USA: A data-driven approach. *Journal of Big Data*, 11(1), 1–24.
[9] Islam, M. R., Nasiruddin, M., Karmakar, M., Akter, R., Khan, M. T., Sayeed, A. A., & Amin, A. (2024). Leveraging Advanced Machine Learning Algorithms for Enhanced Cyberattack Detection on US Business Networks. *Journal of Business and Management Studies*, 6(5), 213–224. https://doi.org/10.32996/jbms.2024.6.5.23.
[10] Jakir, T., et al. (2023). Machine Learning-Powered Financial Fraud Detection: Building Robust Predictive Models for Transactional Security. *Journal of Economics, Finance and Accounting Studies*, 5(5), 161–180. https://doi.org/10.32996/jefas.2023.5.5.16.
[11] Khan, M. A. U. H., Islam, M. D., Ahmed, I., Rabbi, M. M. K., Anonna, F. R., Zeeshan, M. D., ... & Sadnan, G. M. (2025). Secure Energy Transactions Using Blockchain Leveraging AI for Fraud Detection and Energy Market Stability. *arXiv preprint arXiv:2506.19870*. https://doi.org/10.63332/joph.v5i6.2198.

[12] Moustafa, N., & Slay, J. (2015). UNSW-NB15: A comprehensive data set for network intrusion detection systems (UNSW-NB15 network data set). *2015 Military Communications and Information Systems Conference (MilCIS)*, 1–6. https://doi.org/10.1109/MilCIS.2015.7348942.

[13] Rahman, A., Debnath, P., Ahmed, A., Dalim, H. M., Karmakar, M., Sumon, M. F. I., & Khan, M. A. (2024). Machine learning and network analysis for financial crime detection: Mapping and identifying illicit transaction patterns in global black money transactions. *Gulf Journal of Advance Business Research*, 2(6), 250–272. https://doi.org/10.51594/gjabr.v2i6.49.

[14] Rahman, M. S., Hossain, M. S., Rahman, M. K., Islam, M. R., Sumon, M. F. I., Siam, M. A., & Debnath, P. (2025). Enhancing Supply Chain Transparency with Blockchain: A Data-Driven Analysis of Distributed Ledger Applications. *Journal of Business and Management Studies*, 7(3), 59–77. https://doi.org/10.32996/jbms.2025.7.3.7.

[15] Ray, R. K., Sumsuzoha, M., Faisal, M. H., Chowdhury, S. S., Rahman, Z., Hossain, E., ... & Rahman, M. S. (2025). Harnessing Machine Learning and AI to Analyze the Impact of Digital Finance on Urban Economic Resilience in the USA. *Journal of Ecohumanism*, 4(2), 1417–1442. https://doi.org/10.62754/joe.v4i2.6515.

[16] Stiawan, D., Idris, M. Y. I., Heryanto, B., Budiarto, R., & Ab Razak, M. F. (2023). Cyber Threat Landscape and Challenges for Insider Threat Detection: A Systematic Review. *IEEE Access*, 11, 87324–87341.

[17] Sultana, K. S., Begum, M., Abed, J., Siam, M. A., Sadnan, G. A., Shatyi, S. S., & Billah, M. (2025). Blockchain-Based Green Edge Computing: Optimizing Energy Efficiency with Decentralized AI Frameworks. *Journal of Computer Science and Technology Studies*, 7(1), 386–408. https://doi.org/10.32996/jcsts.2025.7.1.29.

[18] Tuor, A., Kaplan, S., Hutchinson, B., Nichols, N., & Robinson, S. (2017). Deep learning for unsupervised insider threat detection in structured cybersecurity data streams. *Proceedings of the AAAI Workshops*, 17(WS-17-01), 103–110.